\documentclass{elsart}

\usepackage{graphicx}
\usepackage{amssymb}

\begin{document}

\begin{frontmatter}

\title{Synchronization of globally coupled nonidentical maps with
inhomogeneous delayed interactions}
\author[] {Arturo C. Mart\'{\i}},
\ead{marti@fisica.edu.uy}
\author[] {C. Masoller}
\ead{cris@fisica.edu.uy}

\address[]{Instituto de F\'{\i}sica, Facultad de Ciencias,
Universidad de la Rep\'ublica, Igu\'a  4225, Montevideo 11400,
Uruguay}

\begin{abstract}
  We study the synchronization of a coupled map lattice consisting of
  a one-dimen\-sional chain of logistic maps. We consider global
  coupling with a time-delay that takes into account the finite
  velocity of propagation of interactions. We recently showed that
  clustering occurs for weak coupling, while for strong coupling the
  array synchronizes into a global state where each element sees all
  other elements in its current, present state [Physica A {\bf 325}
  (2003) 186, Phys. Rev. E {\bf 67} (2003) 056219]. In this paper we
  study the effects of in-homogeneities, both in the individual maps,
  which are non-identical maps evolving in period-2 orbits, and in the
  connection links, which have non-uniform strengths. We find that the
  global synchronization regime occurring for strong coupling is
  robust to heterogeneities: for strong enough average coupling the
  inhomogeneous array still synchronizes in a global state in which
  each element sees the other elements in positions close to its
  current state. However, the clustering behaviour occurring for small
  coupling is sensitive to inhomogeneities and differs from that
  occurring in the homogeneous array.

\end{abstract}

\end{frontmatter}

Globally coupled units with time delays and inhomogeneous interactions
arise in a variety of fields. In the fields of economics and social
sciences, a multi-agent model that allows for temporally distributed
asymmetric interactions between agents was recently proposed in Ref.
\cite{econo}. The model defined a coupled map lattice with
interactions obeying a Gaussian law and transmitted through a gamma
pattern of delays. In the field of optics, a globally coupled laser
array with feedback has been proposed for exploring the complex
dynamics of systems with high connectivity \cite{otsuka,jordi,mandel}.
Globally and locally coupled maps have been employed to study
synchronization \cite{book} in a great variety of fields ranging from
activity patterns in pulse-coupled neuron ensembles
\cite{kurths,haken} to dynamics models of atmospheric circulation
\cite{gallas}.

Since they were introduced by Kaneko, globally coupled logistic maps
have turned out to be a paradigmatic example in the study of
spatiotemporal dynamics in extended chaotic systems \cite{kaneko}.  In
the simplest form all maps are identical and interact via their mean
field with a common coupling:
\begin{equation}
\label{mapa_1}
x_i(t+1)= (1-\varepsilon) f[x_i(t)] + {{\varepsilon}\over{N}} 
\sum_{j=1}^N f[x_j(t)].
\end{equation}
Here $t$ is a discrete time index, $i$ is a discrete spatial index:
$i=1\dots N$ where $N$ is the system size, $f(x)=ax(1-x)$ is the
logistic map, and $\varepsilon$ is the coupling strength.  Even though
the model has only two parameters (the common nonlinearity $a$ and the
coupling strength $\varepsilon$), it exhibits a rich variety of
behaviours. For a large $\epsilon$ the maps synchronize in a global
state, and evolve together as a single logistic map. For intermediate
coupling the array divides into two clusters ($x_i(t)=x_j(t)$ if $i$
and $j$ belong to the same cluster) which oscillate in opposite
phases. For small coupling the number of clusters increases but is
nearly independent of the total number of maps in the system. Finally,
for very small coupling the number of clusters is proportional to $N$
\cite{shimada}.

We have recently studied effect of time-delayed interactions
\cite{marti1,marti2}. We considered the array:
\begin{equation}
\label{mapa0}
x_i(t+1)= (1-\varepsilon) f[x_i(t)] + {{\varepsilon}\over{N}}
\sum_{j=1}^N f[x_j(t-\tau_{ij})]
\end{equation}
where $\tau_{ij}$ is a time delay, proportional to the distance
between the $i$th and $j$th maps. We took $\tau_{ij} = k |i-j|$, where
$k$ is the inverse of the velocity of the interaction signal
traveling along the array.

We found that for weak coupling the array divides into clusters,
and the behavior of the individual maps within each cluster depends on
the delay times. For strong enough coupling, the array synchronizes
into a single cluster (globally synchronized state). In this state the
elements of the array evolve along a orbit of the uncoupled map, while
the spatial correlation along the array is such that an individual map
sees all other maps in his present, current, state:
\begin{equation}
\label{solution}
x_j(t-\tau_{ij})=x_i(t).
\end{equation}

It was also found that for values of the nonlinear parameter $a$ such
that the uncoupled maps are chaotic, time-delayed mutual coupling
suppress the chaotic behavior by stabilizing a periodic orbit which is
unstable for the uncoupled maps.

In this paper we extend the previous study to assess the influence of
heterogeneities. We consider the following array:
\begin{equation}
\label{mapa}
x_i(t+1)= (1-\varepsilon_i) f[a_i,x_i(t)] + {{1}\over{N}}
\sum_{j=1}^N \varepsilon_{i,j}f[a_j,x_j(t-\tau_{ij})]
\end{equation}
where $f(a,x)=ax(1-x)$, $\varepsilon_{i,j}$ is the strength of the
link coupling the maps $i$ and $j$, $\varepsilon_i$ is the average
connection strength of the site $i$:
\begin{equation}
\label{eprom}
\varepsilon_i={{1}\over{N}} \sum_{j=1}^N \varepsilon_{i,j}
\end{equation}
and $\tau_{ij} = k |i-j|$. We perform numerical simulations with the
aim of studying the effects of inhomogeneities in the maps and in the
strength of the coupling links. We consider two different situations:

1) Nonidentical maps coupled with identical coupling strengths
($\epsilon_{i,j}=\epsilon$ $\forall$ $i$, $j$). The value of $a_i$ is
random, uniformly distributed in the interval ($a_0-\delta
a$,$a_0+\delta a$). We limit ourselves to consider maps in the
period-2 region ($a_i\in [3,3.449\dots]$). The study of the
synchronization of the array when the individual maps without coupling
are either in fixed points or in period-$n$ orbits is left for future
work.

2) Identical maps ($a_i=a_0$ $\forall$ $i$) coupled with nonindentical
links.  In this case we consider two different situations: (a) The
value of $\varepsilon_{i,j}$ random, uniformly distributed in the
interval (0,$\epsilon$). (b) $\varepsilon_{i,j}$ varies in interval
[0,$\epsilon$], such that the strength of a link decreases with its
length. We take $\varepsilon_{i,j} =
\epsilon[1-\tau_{i,j}/\max(\tau_{i,j})]$. It can be noticed from Eqs.
(\ref{mapa},\ref{eprom}) that the globally synchronized state
Eq.(\ref{solution}) is also an exact solution of the array when the
maps are identical but the connection strengths are not.

Let us first illustrate the transition to global synchronization in
the case of inhomogeneous maps connected through homogeneous links
($\epsilon_{i,j}=\epsilon$ $\forall$ $i$,$j$). We exemplify results
for $k$ odd. For $k$ odd the globally synchronized state of the
homogeneous array is the anti-phase state:
$x^A_i(t)=x_a,x_b,x_a,\dots$; $x^A_i(t+1)=x_b,x_a,x_b,\dots$. Here
$x_a$ and $x_b$ are the points of the limit cycle of the map
$f(a_0,x)$. Similar results are observed for $k$ even (in this case
the globally synchronized state is the in-phase state:
$x^I_i(t)=x_a,x_a,x_a,\dots$; $x^I_i(t+1)=x_b,x_b,x_b,\dots$).

Figures 1(a) and 1(b) display bifurcation diagrams for increasing
$\epsilon$, which were done in the following way: we chose a random
initial condition [$x_i(0)$ with $i=1,\dots,N$], and plotted a certain
number of time-consecutive values of $x_i(t)$ (with $t$ large enough
and $i=1,\dots,N$) vs. the coupling strength $\epsilon$. Since the
initial condition fixes the cluster partition of the array, the same
initial condition was used for all values of $\epsilon$. For
comparison, Fig. 1(a) displays the bifurcation diagram for a
homogeneous chain ($\delta a=0$), while Fig. 1(b) displays results for
$\delta a=0.2$.  While a sharp transition to global synchronization is
observed for $\delta a=0$, a smoother transition, reminiscent of
bifurcations in the presence of noise, is observed for $\delta
a\not=0$.
To illustrate this transition with more detail we show in Fig. 2 the
configuration of the array for several values of $\epsilon$ (the
values of $\epsilon$ are indicated with an arrow in Fig. 1). It is
observed that the array gradually evolves to the antiphase state. For
low $\epsilon$ there is a large dispersion in the values of $x_i$ and
''defects'' are observed. The dispersion and the number of defects
gradually diminishes as $\epsilon$ increases.
\begin{figure}
  \centerline{\includegraphics[width=13cm,height=5.0cm]{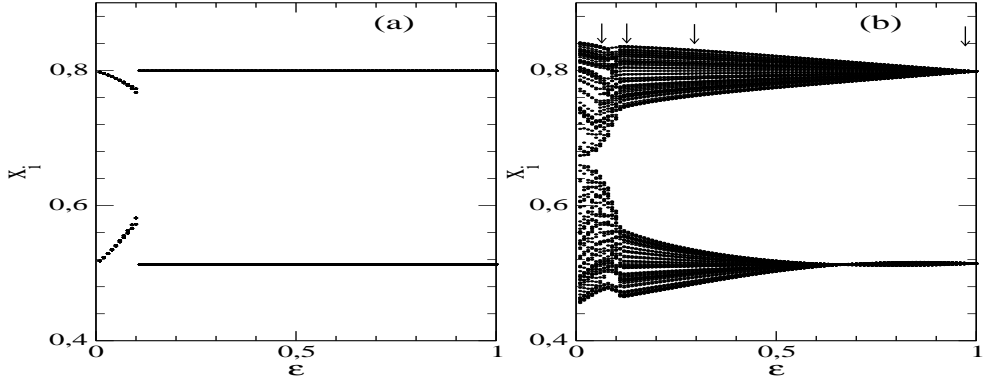}}
\caption{Bifurcation diagram for homogeneous (a) ($\delta a=0.$)
  and inhomogeneous (b) ($\delta a=0.2$) array. Parameter values:
  $a=3.2$, $k=1$, and $N=100$.  }
\label{diag1}
\end{figure}
\begin{figure}
\centerline{\includegraphics[clip=true,width=13cm,height=6cm]{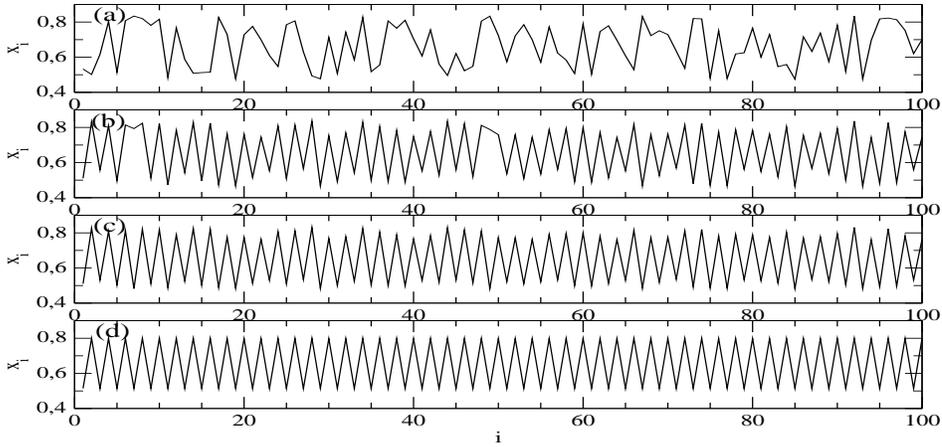}}
\caption{Array configuration at time $t=3000$, for different values
  of the coupling strength $\epsilon$ (a) $\epsilon=0.07$, (b)
  $\epsilon=0.11$, (c) $\epsilon=0.29$, and (d) $\epsilon=0.98$ . Other
  parameters as in Fig. 1 .}
\label{defectos}
\end{figure}

Next we study the case of homogeneous maps ($a_i=a_0$ $\forall$ $i$)
connected through inhomogeneous links. Figure 3 displays bifurcation
diagrams done in the same way as in Fig. 1, i.e., taking the same
initial configuration of the array for all values of $\epsilon$.
However, in this case the links have different coupling strengths
[$\varepsilon_{i,j}$ varies in (0,$\varepsilon$), is either uniformly
distributed or decreasing with $\tau_{i,j}$]. Therefore, in the
horizontal axis we plotted the {\it average} coupling strength
$<\varepsilon>=1/N^2 \sum_i \sum_j \varepsilon_{i,j}$. It can be
observed that as $<\varepsilon>$ increases there is a sharp transition
to global synchronization, both when $\epsilon_{i,j}$ is randomly
distributed in (0,$\epsilon$), and when $\epsilon_{i,j}$ is in
(0,$\epsilon$) decreasing with distance. However, there is different
clustering behaviour, as illustrated in Fig. 4. Figures 4(a) and (b)
display the clustering behaviour for $k=1$ (we point out that for
$k=1$ and $<\varepsilon>$ large enough the array synchronizes in
anti-phase), and Figs. 4(c) and (d) display the clustering behaviour
for $k=2$ (for $k=2$ and $<\varepsilon>$ large enough the array
synchronizes in-phase). For comparison, Figs. 4(a) and 4(c) display
the clustering behaviour of the homogeneous array. In these figures
$\epsilon_{i,j}=\epsilon_{0}$ with $\epsilon_{0}$ equal to the
average coupling strength in Figs. 4(b) and 4(d). In spite of the fact
that the average connection strength is the same, the partition of the
array is different when the connection links have non-uniform
strengths.
\begin{figure}
\centerline{\includegraphics[width=14cm]{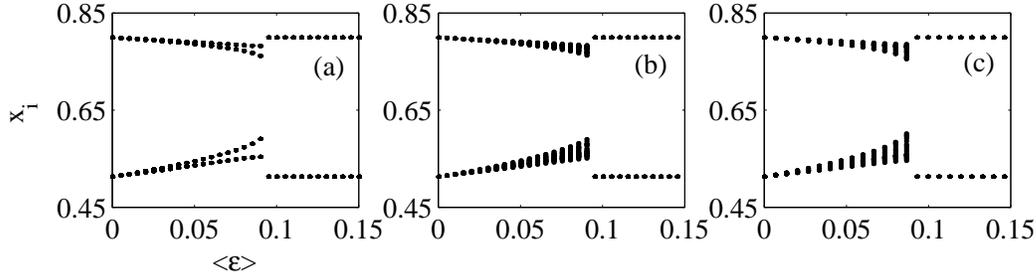}}
\caption{Bifurcation diagram displaying the transition to the globally
  synchronized state. $N=100$, $a_i=3.2$ $\forall$ $i$, and $k=1$. (a)
  $\varepsilon_{i,j}=\varepsilon_0$ $\forall$ $i$ and $j$; (b)
  $\varepsilon_{i,j}$ is uniformly distributed in $(0,\epsilon$); (c)
  $\varepsilon_{i,j}$ decreases with distance.}
\label{f1}
\end{figure}
\begin{figure}
\centerline{\includegraphics[height=5cm]{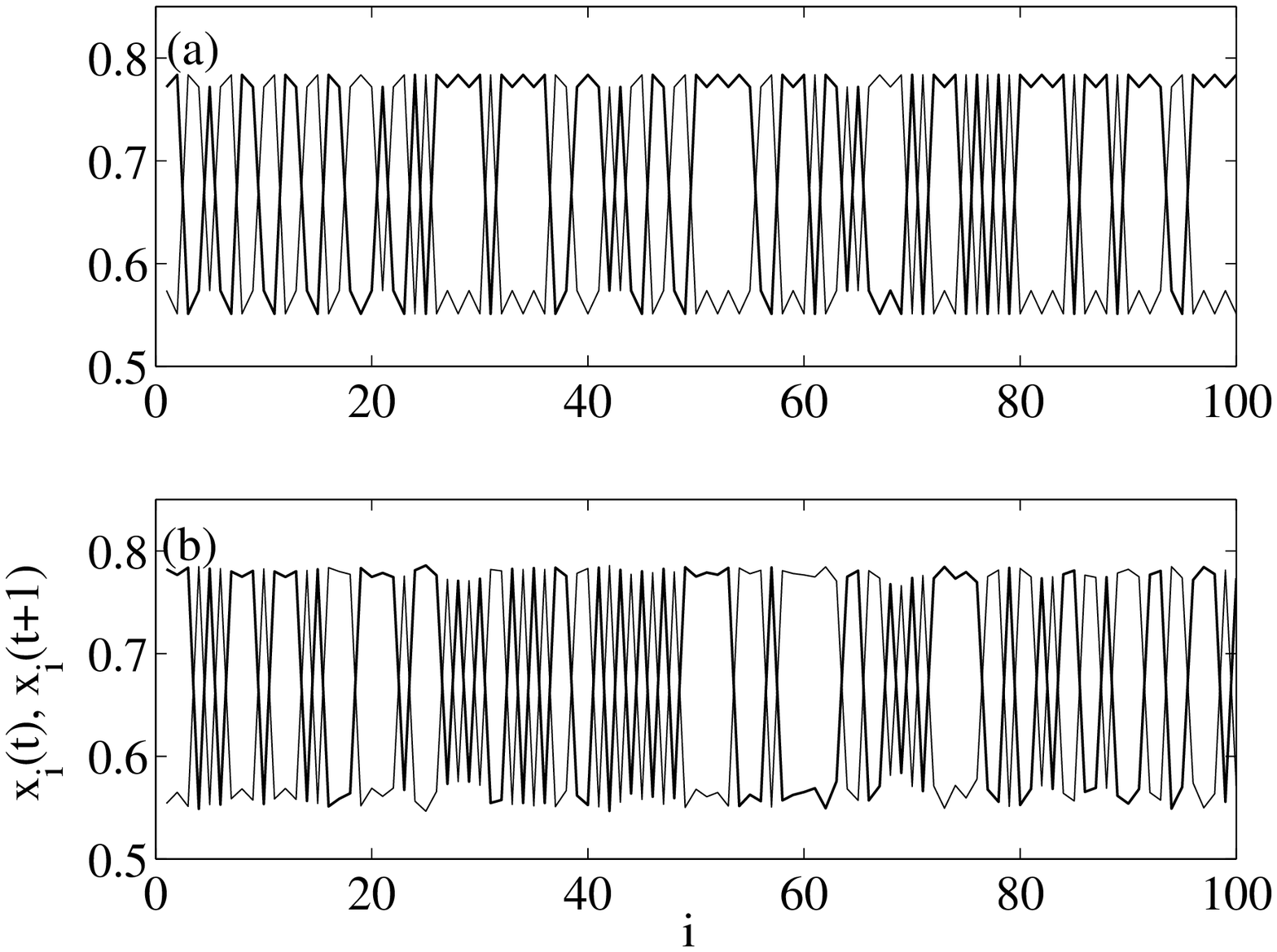}
\includegraphics[height=5cm]{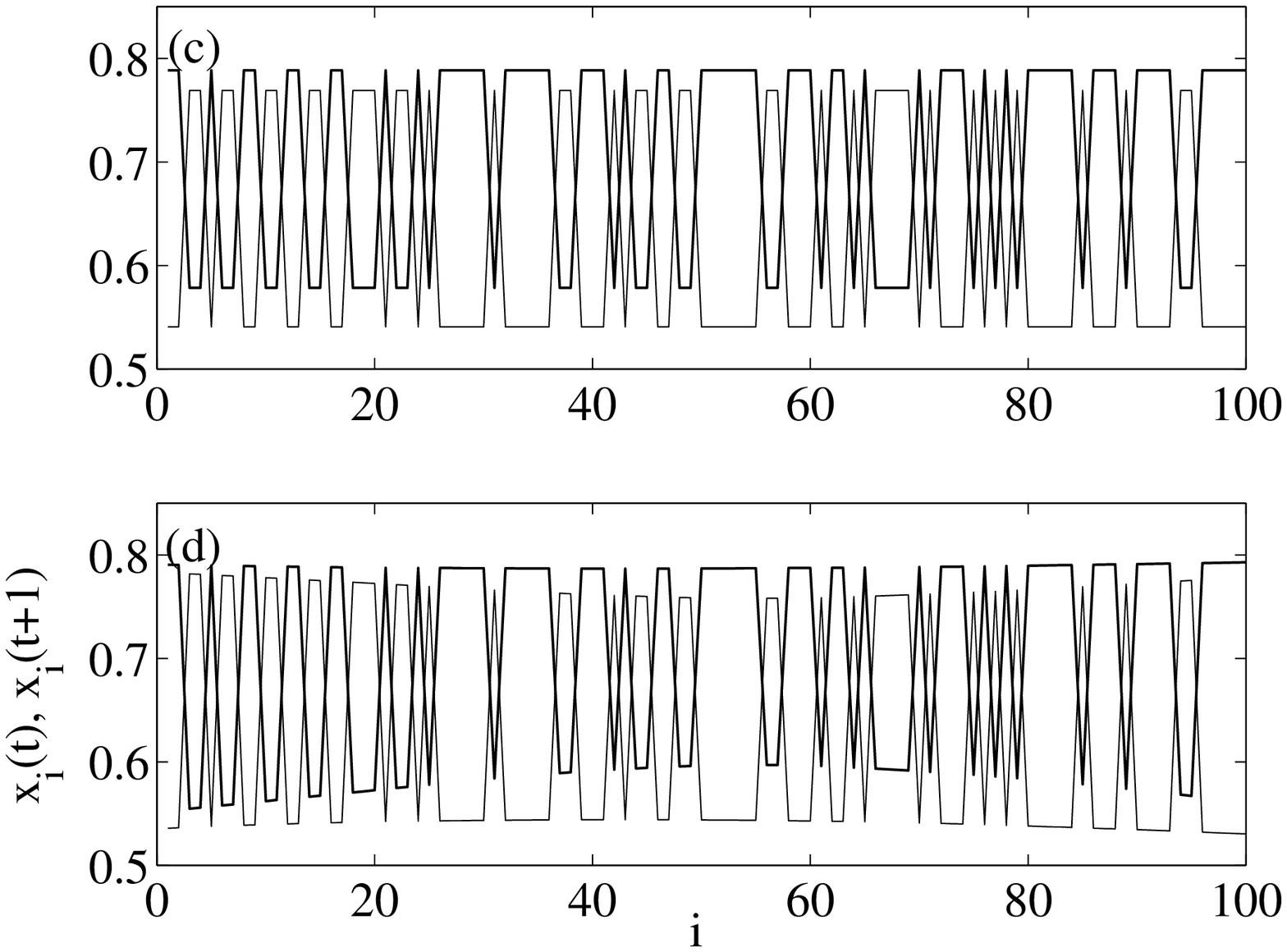}}
\caption{Clustering behaviour for weak coupling strength.
  The thin line shows the array configuration at time $t$, and the
  thick line shows the array configuration at time $t+1$.  $N=100$ and
  $a_i=3.2$ $\forall$ $i$. (a) $k=1$ and $\varepsilon_{i,j}=0.08$
  $\forall$ $i$,$j$; (b) $k=1$ and $\varepsilon_{i,j}$ uniformly
  distributed in [0,0.16], $<\varepsilon_{i,j}>=0.08$; (c) $k=2$ and
  $\varepsilon_{i,j}=0.073$ $\forall$ $i$,$j$; (d) $k=2$ and
  $\epsilon_{i,j} \in (0,0.11)$ decreases with distance,
  $<\varepsilon_{i,j}>=0.073$.}
\label{f2}
\end{figure}

To quantify the effect of in-homogeneities we introduce the following
quantity
\begin{equation}
\sigma (t) = \sqrt {{{1}\over{N}}\sum_i[x_i(t)-x^R_i(t)]^2},
\end{equation}
that measures the deviation of the array configuration, $x_i(t)$, from
a given reference state, $x^R_i(t)$.  $x_i(t)$ is evaluated at time
$t$ with $t$ long enough to let transients die away. The reference
state is the anti-phase state for $k$ odd
($x^R_i(t)=x_a,x_b,x_a,\dots$; $x^R_i(t+1)=x_b,x_a,x_b,\dots$) and the
in-phase state for $k$ even ($x^R_i(t)=x_a,x_a,x_a,\dots$;
$x^R_i(t+1)=x_b,x_b,x_b,\dots$). Here $x_a$ and $x_b$ the points of the
limit cycle of the map $f(a_0,x)$.

\begin{figure}
\centerline{\includegraphics[height=5cm]{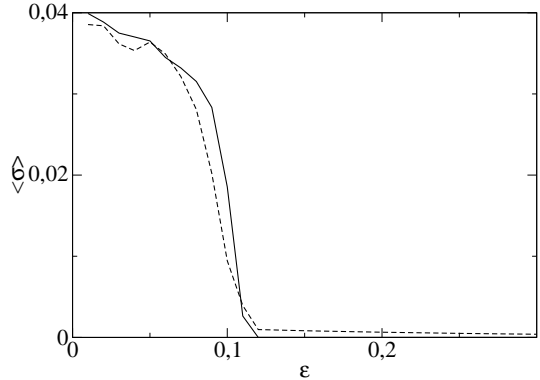}}
\caption{Deviation of the array configuration  vs
  coupling strength averaged over 100 realizations for the cases with
  (continuous line, $\delta a=0.2$) and without dispersion (dashed
  line, $\delta a=0$).  (Parameter values: $N=100$, $k=1$, $a=3.2$)}
\label{sigma-eps}
\end{figure}

Figure \ref{sigma-eps} shows $<\sigma>$ vs. $\epsilon$ where $<\cdot>$
represents an average over different initial configurations $x_i(0)$.
We observe here that in the homogeneous array (solid line) the decay
of the deviation $\sigma$ is more abrupt that in the heterogeneous
array (dashed line).  This is due to related to the sharp transition
to global synchronization shown in Fig.~\ref{diag1}.

To summarize, we have studied the effect of the inhomogeneities in a
linear chain of globally coupled logistic maps with time-delayled
interactions. The inhomogeneities occurs both in the individual maps
which are non-identical and in the connection link which have
non-uniform strengths. We considered the case in which the maps,
without coupling, evolve in a limit cycle of period $P=2$. We found
that the global synchronization regime occurring for strong coupling
is robust to heterogeneities: for strong enough average coupling the
inhomogeneous array still synchronizes in a global state in which each
element sees the other elements in positions close to its current
state. However, the clustering behaviour occurring for small coupling
is sensitive to inhomogeneities and differs from that occurring in the
homogeneous array.

\end{document}